\documentclass{article}

\usepackage{arxiv}

\usepackage[utf8]{inputenc} 
\usepackage[T1]{fontenc}    
\usepackage{hyperref}       
\usepackage{url}            
\usepackage{booktabs}       
\usepackage{amsfonts}       
\usepackage{nicefrac}       
\usepackage{microtype}      
\usepackage{tabularx}
\usepackage{epsfig}
\usepackage{latexsym}
\usepackage{amsmath}
\usepackage{amssymb}
\usepackage{amsfonts}
\usepackage{graphicx}
\usepackage{wrapfig}

\title{Social Distance Characterization by means of Pedestrian Simulation}

\author{Daniel R. Parisi (1)\\
  \texttt{dparisi@itba.edu.ar} \\
   \And
  Germ\'an A. Patterson (1)\\
\And
Lucio Pagni (2) \\
\And
Agustina Osimani (2) \\
\And
Tomas Bacigalupo (2) \\
\And
Juan Godfrid (2) \\
\And
Federico M. Bergagna (2) \\
\And
Manuel Rodriguez Brizi (2) \\
\And
Pedro Momesso (2) \\
\And
Fermin L. Gomez (2) \\
\And
Jimena Lozano (2) \\
\And
Juan Martin Baader (2) \\
\And
Ignacio Ribas (2) \\
\And
Facundo P. Astiz Meyer (2) \\
\And
Miguel Di Luca (2) \\
\And
Nicol\'as E. Barrera (2) \\
\And
Ezequiel M. Keimel \'Alvarez (2) \\
\And
Maite M. Herran Oyhanarte (2) \\
\And
Pedro R. Pingarilho (2) \\
\And
Ximena Zuberbuhler (2) \\
\And
Felipe Gorostiaga (2) \\
\And
\\
(1) Instituto Tecnol\'ogico de Buenos Aires (ITBA) CONICET, 
Lavard\'en 315 (1437), C.A. de Buenos Aires, Argentina.\\
(2) Instituto Tecnol\'ogico de Buenos Aires (ITBA), 
Lavard\'en 315 (1437), C.A. de Buenos Aires, Argentina.\\}

\begin{document}
\maketitle

\begin{abstract}

In the present work, we study how the number of simulated clients (occupancy) affects the social distance in an ideal supermarket. For this, we account for realistic typical dimensions and process time (picking products and checkout). From the simulated trajectories, we measure events of social distance less than 2 m and its duration. Between other observables, we define a social distance coefficient that informs how many events (of a given duration) suffer each agent in the system. These kinds of outputs could be useful for building procedures and protocols in the context of a pandemic allowing to keep low health risks while setting a maximum operating capacity.

\end{abstract}


\section{Introduction}
\label{sec:Int}

One of the measures widely applied in order to mitigate Coronavirus disease (COVID-19) outbreak is to keep a social distance between people \cite{who2020coronavirus}. This distance would act as a physical barrier for small liquid droplets released from the nose or mouth of a potentially infected person. When another person is too close, he/she can breathe in the droplets and get infected. Although COVID-19 is our current concern, the social distance is useful for any contagious disease.

 Of course, there can be other ways of transmission in which the social distance does not work, for example, by touching a surface and then come into contact with any mucous membrane, or if the pedestrian system is indoor without proper ventilation. Also, the social distance required may depend on other preventing measures such as wearing or not a mask. 
 
 Recent works \cite{rathinakumar2020microscopic, harweg2020agent} propose to combine microscopic agent simulation with general disease-transmission mechanisms. However, because of the complexity and uncertainties in the actual knowledge for quantifying these transmission processes, we will not consider in this work any particular contagion mechanism, instead, we will focus on studying the distance between people in an everyday pedestrian facility as an isolated aspect to be integrated in the future by experts considering all mechanism for any particular disease propagation. An antecedent of this kind of analysis was recently reported considering field data from a train station \cite{pouw2020monitoring}.
 
 One of the key questions we try to answer is how to describe the realized social distance for a given occupation of an establishment. For solving this problem, it is necessary to consider the pedestrian displacements and trajectories that they do while performing certain tasks, and thus, a natural tool is to use pedestrian simulation. The time evolution of positions from simulated agents can provide not only the relative distance between agents, but also the duration of events in which the recommended social distance is not kept.
 
 Many industries and shops were closed in different phases of the COVID-19 pandemic. However, grocery shops had to be kept open, in particular supermarkets. To avoid crowding and to keep some physical distance between clients, the authorities lowered the allowed capacity. Different countries regulation adopt social distance requirements between 1 and 2 m \cite{pouw2020monitoring}. In the present study we will consider a distance of 2 m as the social distance threshold, because this distance is an upper bound for the saliva droplets from a human cough which cannot travel more than 2 m in space at approximately zero wind speed \cite{dbouk2020coughing}.
 
 We propose to investigate how the value of the allowed capacity affects the social distance in an ideal supermarket of $448 ~ \text{m}^2$. The results should not be directly extrapolated to other supermarkets or facilities, nevertheless, the methodology and indicators can be useful for applying them to other source of data, being from simulations or field data of other pedestrian systems.

\section{Models}
\label{sec:Mod}

In order to simulate the complex environment and the agent's behavior, the proposed model involves three levels of complexity: operational, tactical, and strategic \cite{hoogendoorn2004pedestrian}.

\subsection{Strategic Level}
\label{subsec:Strategic}

The more general level of the model is in charge of providing a master plan for the agent when it is created. In practical terms for the present system, it gives a list of $n_p$ products for agents to  acquire (a shopping list). Each one of the $n_p$ items is randomly chosen between a total of $m_p$ products available. Also, they are identified with a unique target location ($\mathbf{x}_{pn}$) in the supermarket.   

Once the agent is initialized with its shopping list, the strategic level shows the first item in the list to the agent. The agent will move toward it using the lower levels of the model. When the agent reaches the position of the product, it will spend a picking time ($t_p$) choosing and taking the product, after which the strategic level will provide to the agent the next item on the list. 

After the list of products is completed, the agent must proceed to the less busy supermarket checkout line. It will adopt a queuing behavior until it gets the checkout desk and spends a time $t_{co}$ processing its purchase. 

\subsection{Tactical Level}
\label{subsec:Tactical}

The function of the tactical level is to provide successive visible targets to the agent that will guide it to the location of the desired product ($\mathbf{x}_{pn}$) or checkout line. As inputs, the tactical module takes the current agent position ($\mathbf{x}_i(t)$) and the position of the current product ($\mathbf{x}_{pn}$) in the list. The output is a temporal target ($\mathbf{x}_{v}(t)$) visible from the current position of the agent. The definition of visibility is that if we take a virtual segment between ($\mathbf{x}_{i}(t)$) and ($\mathbf{x}_{v}(t)$), this segment does not intersect any of the walls or obstacles (shelves).

The information delivered by the tactical module is obtained by implementing a squared network connecting all the accessible areas of the simulated layout (see Fig.\ref{Fig:Blueprint} ). Given any two points in the walkable domain, the corresponding nearest points on the network are found and then the shorter path between these points is computed by using the A* algorithm \cite{yao2010path}.

Once the path on the network is defined, the temporary target $\mathbf{x}_{v}(t)$ is chosen as the farthest visible point on that path seen from the current agent position. Clearly, $\mathbf{x}_{v}(t)$ will change with time, as the position of the agent changes. When the product target is visible from the agent, this is set as the visible target and the network path is no longer considered until a new product should be found.

\subsection{Operational Level}
\label{subsec:Operational}

As the lower level describing the short-range movements of agents we propose an extended version of the ``Contractile Particle Model'' \cite{baglietto2011continuous}. This extension will provide an efficient navigation allowing to avoid potential collisions with other agents and obstacles. 

The basic model is a first-order model in which particles have continuous variable radii, positions and velocities that change following certain rules. Specifically, the position is updated as

\begin{equation}
\label{eq:foe}
\mathbf{x}^i(t + \Delta t) = \mathbf{x}^i(t) + \mathbf{v}^i  \Delta t\ ,
\end{equation}
\vspace{0 pt}

where $\mathbf{v}^i$ is the desired velocity and $\mathbf{x}^i(t)$ the position at time $t$. The radius of the $i^{th}$ particle ($r^i$) is dynamically adjusted between $r^i_{min}$ and $r^i_{max}$. When this radius has large values, it represents a personal distance necessary for taking steps, but when it has low values, it represents a hard incompressible nucleus which limits maximum densities. 

When particles are not in contact, the desired velocity $\mathbf {v}^i$ points toward the visible target with a magnitude proportional to its radius,

\begin{equation}
\mathbf{v}^i = \mathbf{e}_t^i ~ v\ ,
\label{eqvi}
\end{equation}
\vspace{0 pt}

where the direction $\mathbf{e}_t^i$ and the magnitude $v$ are defined by the following equations
\begin{equation}
\mathbf{e}_t^i = \frac {(\mathbf{X}_v - \mathbf{X}^i)}{|(\mathbf{X}_v - \mathbf{X}^i)|}\ ,
\label{eqetvi}
\end{equation}

\begin{equation}
v = v_{d}  [\frac {(r-r_{min})} {(r_{max}-r_{min})} ]\ ,
\label{eqmvi}
\end{equation}
where $v_{d}$ is the desired speed.

While the radius has not reached the maximum $r_{max}$, it increases in each time step following 
\begin{equation}
\Delta r= \frac {r_{max}} { (\frac {\tau} {\Delta t}) }.
\label{}
\end{equation}
being $\tau$ a characteristic time in which the agent reaches its desired speed as if it was free, and $\Delta t$ is the simulation time step of Eq. \ref{eq:foe}.
When two particles enter into contact ($d_{ij}~=~|\mathbf {x}^i-\mathbf {x}^j| - (r_i +r_j) ~ < ~ 0$) both radius instantaneously collapse to the minimum values while an escape velocity appears moving the particles in the directions that will separate the overlap:

\begin{equation}
\mathbf{e}^{ij} = \frac{(\mathbf{x}^i - \mathbf{x}^j)}{ | \mathbf{x}^i - \mathbf{x}^j |}.
\label{eq:6}
\end{equation}

The escape velocity has the magnitude of the free speed and, thus, it can be written as $\mathbf{v}_e^i = v_{d}~\mathbf{e}^{ij}$. This velocity is only applied during one simulation step because, as the radii are simultaneously collapsing, the agents are no longer overlapping.

Up to here, we described the basic CPM as it appears in Ref. \cite{baglietto2011continuous}. This model satisfactorily described experimental data of specific flow rates and fundamental diagrams of pedestrian dynamics. However, particles do not anticipate any collisions, and this capacity is a fundamental requirement for simulating the ideal supermarket (displaying low and medium densities, and agents circulating in different directions). For this, we propose to extend the calculation of the agent velocity (Eq. \ref{eqvi}) by considering a simple avoidance mechanism.

The general idea is that the self-propelled particle will produce any action only through changing its desired velocity $\mathbf{v}_i(t)$ as stated in Ref. \cite{martin2020pedestrian}. In this case, the new mechanism will only change the direction of desired velocity $\mathbf{v}$ depending on the neighbor particles and obstacles. First, the collision vector ($\mathbf{n_c}^{i}$) is calculated
\begin{equation}
\mathbf{n_c}^{i} = \mathbf{e}^{ij} ~ A_p ~ e^{-d_{ij}/B_p} ~ cos(\theta_j) + \mathbf{e}^{ik} ~ A_w ~ e^{-d_{ij}/B_w} ~ cos(\theta_k) + \hat{\eta}
\label{eq:na}
\end{equation}
where $j$ indicates the nearest visible neighbor, $k$ the nearest point of the nearest visible wall or obstacle, and $\hat{\eta}$ is a noise term for breaking possible symmetric situations.

Then the avoidance direction is obtained from
\begin{equation}
\mathbf{e}^i_a = \frac{(\mathbf{n_c}^{i} + \mathbf{e}_t^i )}{|(\mathbf{n_c}^{i} + \mathbf{e}_t^i )|}
\end{equation}

and finally the velocity of the particle to be used in Eq. (\ref{eq:foe}), if particles are not in contact, is 
\begin{equation}
\mathbf{v}^{i} ~ = v ~ \mathbf{e}^i_a.
\label{eq:newCPMvel}
\end{equation}

\begin{figure}
    \centering
    \centerline{\includegraphics[width=0.9\textwidth]{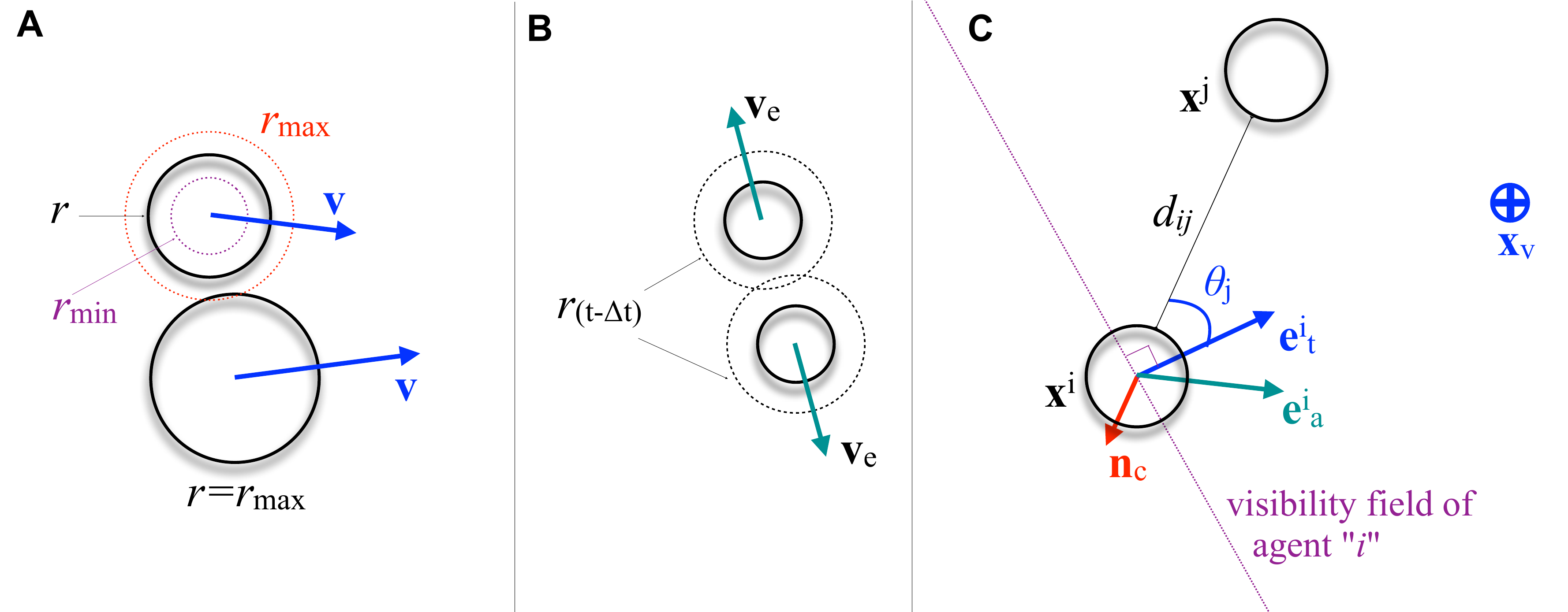}}
    \caption{Contractile particle model. A: Two particles without contact. B: Two particles having had an overlap in the previous time step (dash circles) collapse their radii and take the escape velocity. A and B correspond to the original CPM. C: Modification considering an avoidance direction.}
  \label{Fig:CPM}
\end{figure}

In Fig.\ref{Fig:CPM} the vectors associated with the original and modified model can be seen in detail.

\subsubsection{States of agents}
\label{subsec:States}

Because the agents must perform different tasks, more complex than just going from one point to another, it was necessary to define five behavioral states. This was achieved by setting different model parameters and movement patterns. More concisely, the five behavioral states of agents were:

- \emph{Going}: This is the normal walking behavior when going from one arbitrary point to another with the standard velocity and model parameters. Only in this state, the agent uses the modified CPM velocity (Eq. \ref{eq:newCPMvel}) for avoiding potential collisions.

The rest of the behavioral states only use the basic CPM (Eq. \ref{eq:foe} to \ref{eq:6}).

- \emph{Approximating}: When the agent is closer than 2 m of the current product, it reduces its desired speed and, because of how parameters are set, it will not be forced to reach it if there is another agent buying a product in the same target $\mathbf{x}_{pn}$.

- \emph{Picking}: Once the agent reaches the product (closer than 0.1 m) a timer starts and it will remain in the same position (Eq. \ref{eq:foe} does not update its position) until the picking time ($t_p$) is up.

- \emph{Leaving}: After spending the time $t_p$, the agent leaves the current location going to the next product on the list. While abandoning this position, it could find other waiting agents (in Approximating behavioral state) and, thus, its parameters must be such that it can make its way through. Once the agent is farther than 2 m from the past product, it changes to the "going" behavioral state.

- \emph{Queuing}: Finally, when the agent completes the shopping list, it proceeds to the checkout desks by choosing the one with the smaller line. It waits at a distance of 1.5 m from the previous queuing agent, and when it reaches the checkout position, it remains there for a $t_{co}$ time.

By considering these behavioral states in the agent model, the conflicts and deadlock situations are minimized. In this way, this model improvement allows us to simulate higher densities than with the basic operational models.

Finally, as a closure for this general section (\ref{sec:Mod}) about the model, we point out that for the sake of comparison, we also implement other operational models: the "Social Force Model" \cite{helbing2000simulating, johansson2007specification}, and the "Predictive Collision avoidance Model" (PCA) \cite{karamouzas2009predictive}. The results for all models will be compared in selected observables, while the more deep study is performed using the modified CPM described above.

\section{Simulations}
\label{sec:Sim}

\begin{figure}
    \centering
    \centerline{\includegraphics[width=0.9\textwidth]{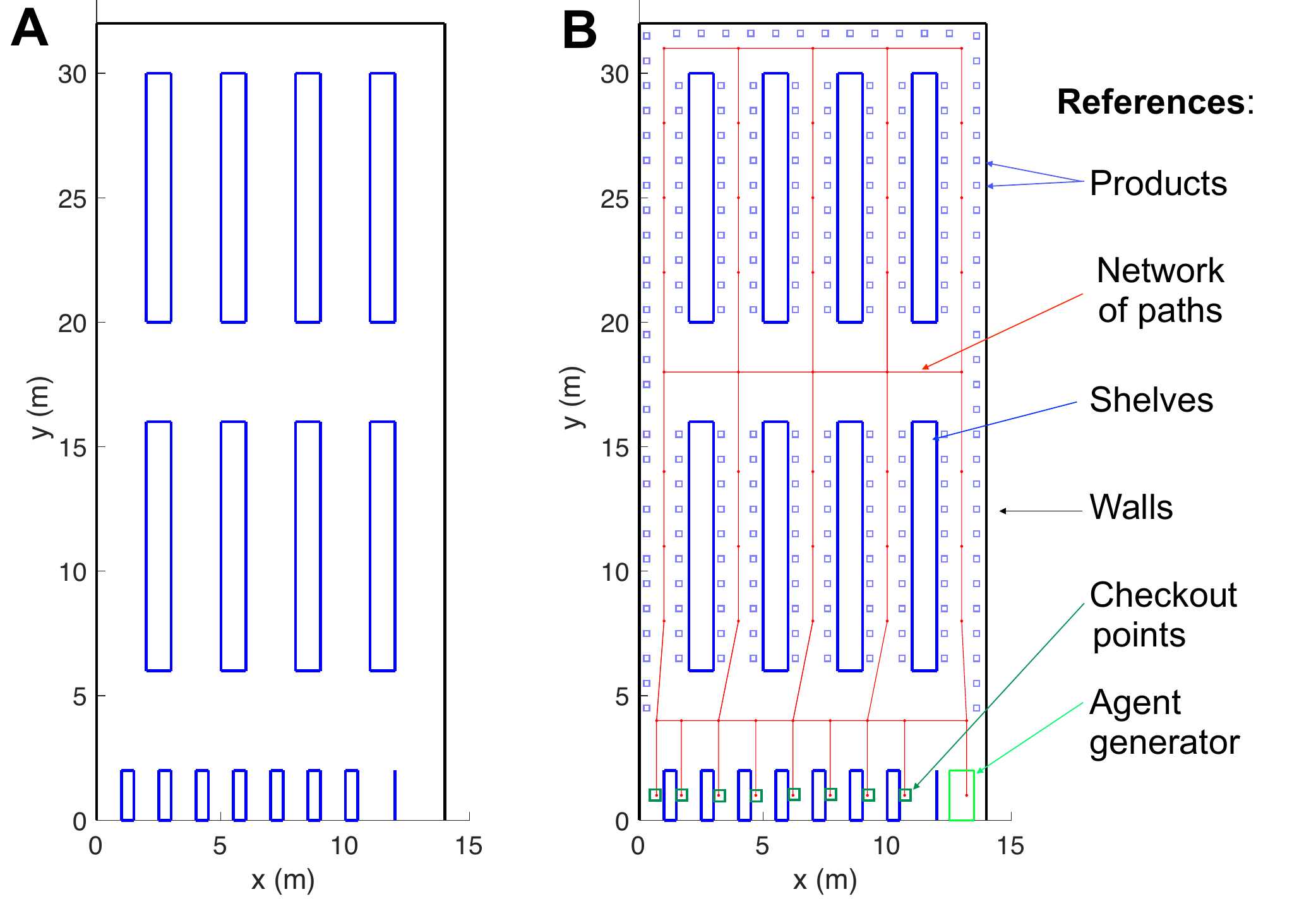}}
    \caption{The ideal supermarket layout. A: Only walls and obstacles. B: The rest of the component of the model as described in Sec.\ref{sec:Mod}.}
  \label{Fig:Blueprint}
\end{figure}

The plant of 448 $\text{m}^2$ of the ideal supermarket to be simulated is shown in Fig. \ref{Fig:Blueprint}. The dimension of shelves (1 m x 10 m) and corridor width (2 m) are taken from typical real systems. Also, the different process time and other data considered were provided by a supermarket chain from Argentina.

We define $N$ as the allowed capacity or the occupation of the supermarket as the total number of agents buying simultaneously inside the system. This is the more important input to be varied in our study and it ranges from $N=2$ to $N=92$.

The agent generator produces an inflow of 1 agent every 5 seconds until it reaches the $N$ value for the simulation. From that moment on, the agent generator monitors the occupation, and it generates a new agent every time an existing agent completes its tasks and is removed from the simulation. By doing this, the value of $N$ is maintained constant over the whole simulation.

Every agent created by the generator is equipped with a shopping list of exactly $n_p ~ = ~ 15 $ items, which are chosen randomly from a total of 228 items available (shown in Fig.\ref{Fig:Blueprint} B). The corresponding product locations ($\mathbf{x}_{pn}$) are separated by one meter between adjacent locations. Agents visiting the products on its lists, spend a picking time with a uniform distribution ($t_p~\in~[60\text{s},90\text{s}]$). After completing the lists, agents choose the shortest queue to one of the eight checkout points shown in Fig. (\ref{Fig:Blueprint} B). The ideal supermarket has a maximum of four queues, and each of them leads to two checkout desks. The distance between the agents that form the queue is 1.5 m. The first positions on these queues are at a distance of 3 m (at $y~=~4$ m, in Fig.\ref{Fig:Blueprint}) from the checkout points. Once an agent reaches the cashier (at $y~=~1$ m, in Fig.\ref{Fig:Blueprint}) it spends a checkout time $t_{co}$ uniformly distribute between $t_{co} \in$ [120s, 240s].

For each value of $N$, we simulated two hours (7200 s) and recorded the state of the system every $\Delta t2 ~ = ~ 0.5$ s, thus producing 14400 data files with positions, velocity, and behavioral state of agents.

The simulation time step $\Delta t$ used in Eq.(\ref{eq:foe}) was for all simulations $\Delta t~ = ~ 0.05$ s.

The noise term in Eq. (\ref{eq:na}) is a random vector, which components $\eta_x$ and $\eta_y$ are uniformly distributed between the range $\eta_x~=~\eta_y~=~[-0.1~\text{m/s}, 0.1~\text{m/s}]$. And the relaxation time $\tau$ is set to $\tau ~ = ~ 0.5 $ s.  

The rest of the model parameters depends on the behavioral state of the agent. For the case of "going", the parameters of the avoidance mechanism described in Eq.(\ref{eq:na}) are: $A_a ~ = ~ 1.25 $, $B_a ~ = ~  1.25 $ m, $A_w ~ = ~ 15 $ and $B_w ~ = ~ 0.15 $ m.

The other behavioral states only implement the original CPM (without avoidance mechanism) with the parameters displayed in Table \ref{table:1}.

\begin{table}[h!]
\centering
\begin{tabularx}{1\textwidth} { 
  | >{\raggedright\arraybackslash}X 
  | >{\centering\arraybackslash}X 
  | >{\centering\arraybackslash}X 
  | >{\centering\arraybackslash}X 
  | >{\centering\arraybackslash}X 
  | >{\centering\arraybackslash}X | }
 \hline
 Behavioral State  &  Going & Approximating & Picking & Leaving & Queuing \\
 \hline
 $r_{min}$ (m) & 0.1   & 0.1  & 0.2 & 0.1 & 0.1 \\
 \hline
 $r_{max}$ (m) & 0.37 & 0.35  & 0.2 & 0.3 & 0.12 \\
 \hline
  $v_d$ (m/s) & 0.7  & 0.5   & 0   & 0.9 & 0 or 0.5 \\
 \hline
\end{tabularx}
\caption{Parameters of the CPM operational model for all the behavioral states.}
\label{table:1}
\end{table}

\section{Results}
\label{sec:Res}

\subsection{General Aspects}
\label{Subsec:General}

\begin{figure}
    \centering
    \centerline{\includegraphics[width=1\textwidth]{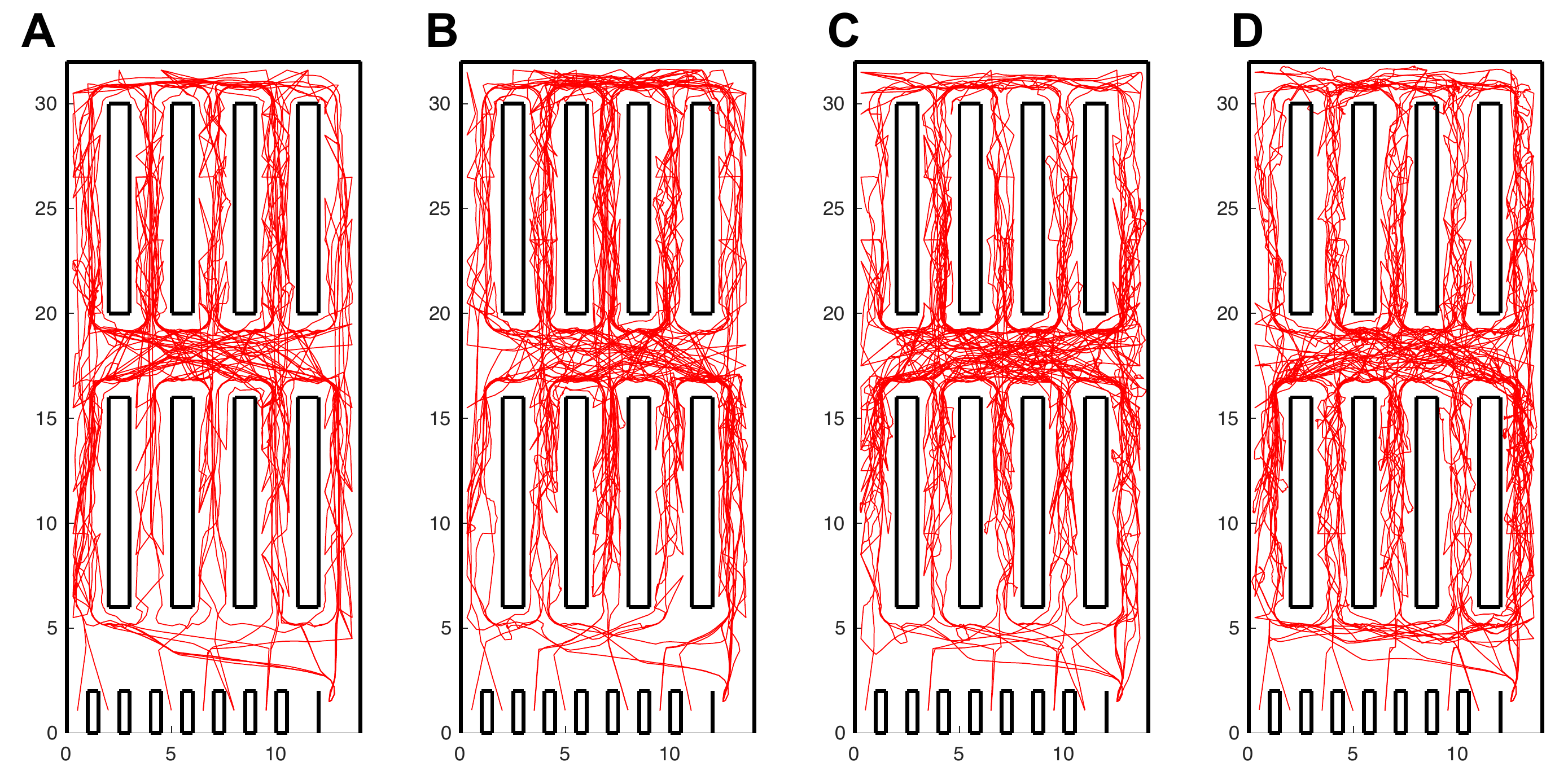}}
    \caption{Trajectories. Ten random trajectories where chosen at simulation near 1 hour for different occupancies. A: $N~=~14$, B: $N~=~35$, C: $N~=~62$, D: $N~=~92$.}
  \label{Fig:Trayect}
\end{figure}

\begin{figure}
    \centering
    \centerline{\includegraphics[width=1\textwidth]{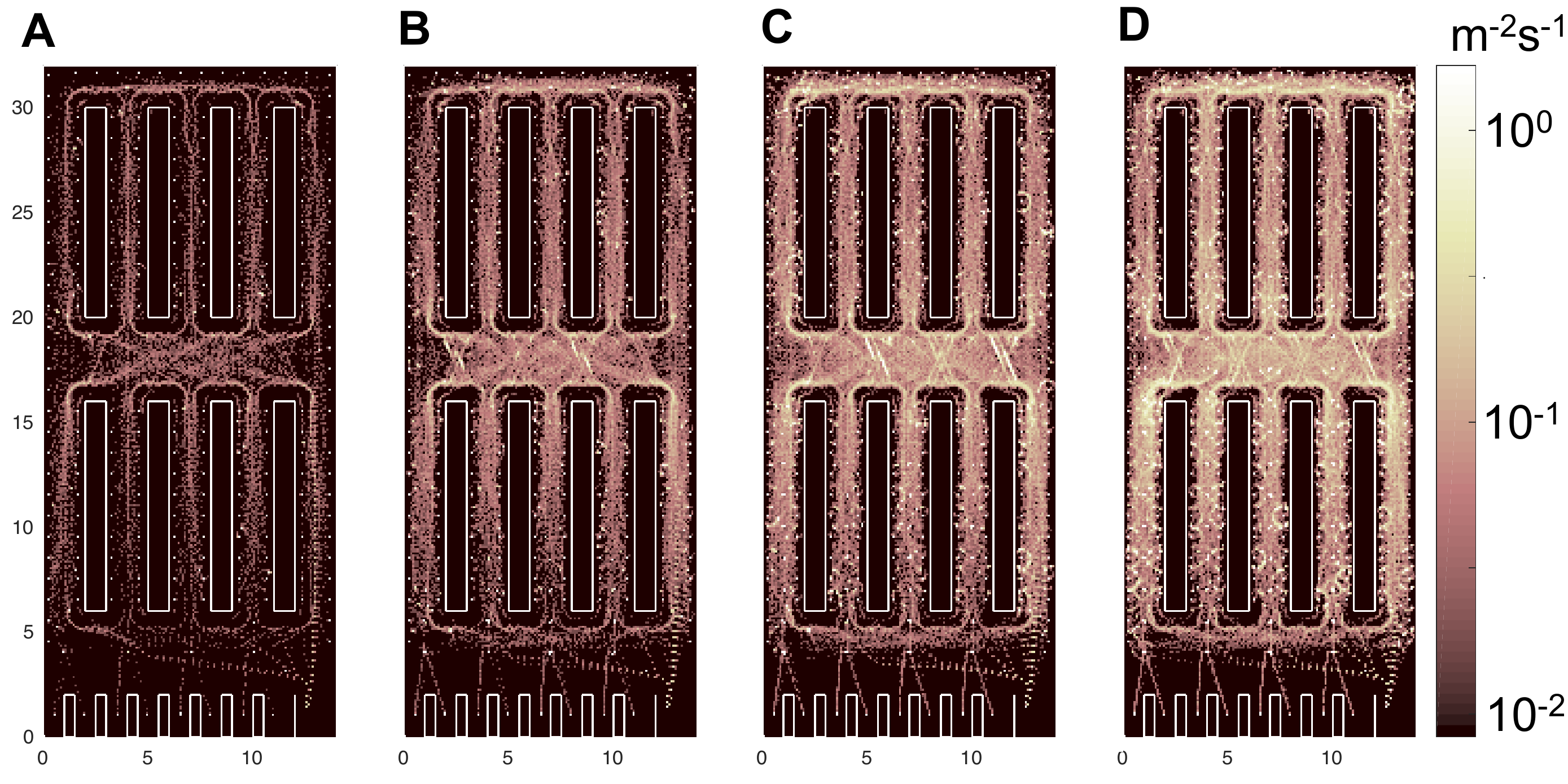}}
    \caption{Density maps averaged during the 2 hours simulation time for different occupancies. A: $N~=~14$, B: $N~=~35$, C: $N~=~62$, D: $N~=~92$.}
  \label{Fig:HeatMap}
\end{figure}

We first show general results of the simulated supermarket by displaying typical trajectories (Fig. \ref{Fig:Trayect}) and density fields (Fig. \ref{Fig:HeatMap}).  

Figure \ref{Fig:Trayect} plots ten random chosen trajectories in the second hour of simulations for the selected $N$ values. Qualitatively it can be seen more intricate trajectory patterns as the occupancy increases. However, in all cases, it can be observed that the available area is uniformly visited by simulated agents while visiting their product list. Complementary information is shown in Fig.\ref{Fig:HeatMap}, where the density is averaged over all the simulation time (2 hours). As expected, greater occupancy presents greater mean density values. Besides, these density fields present higher values at the spots that agents stay longer, revealing picking points of products and predefined places at queuing.

\begin{figure}
    \centering
    \centerline{\includegraphics[width=1\textwidth]{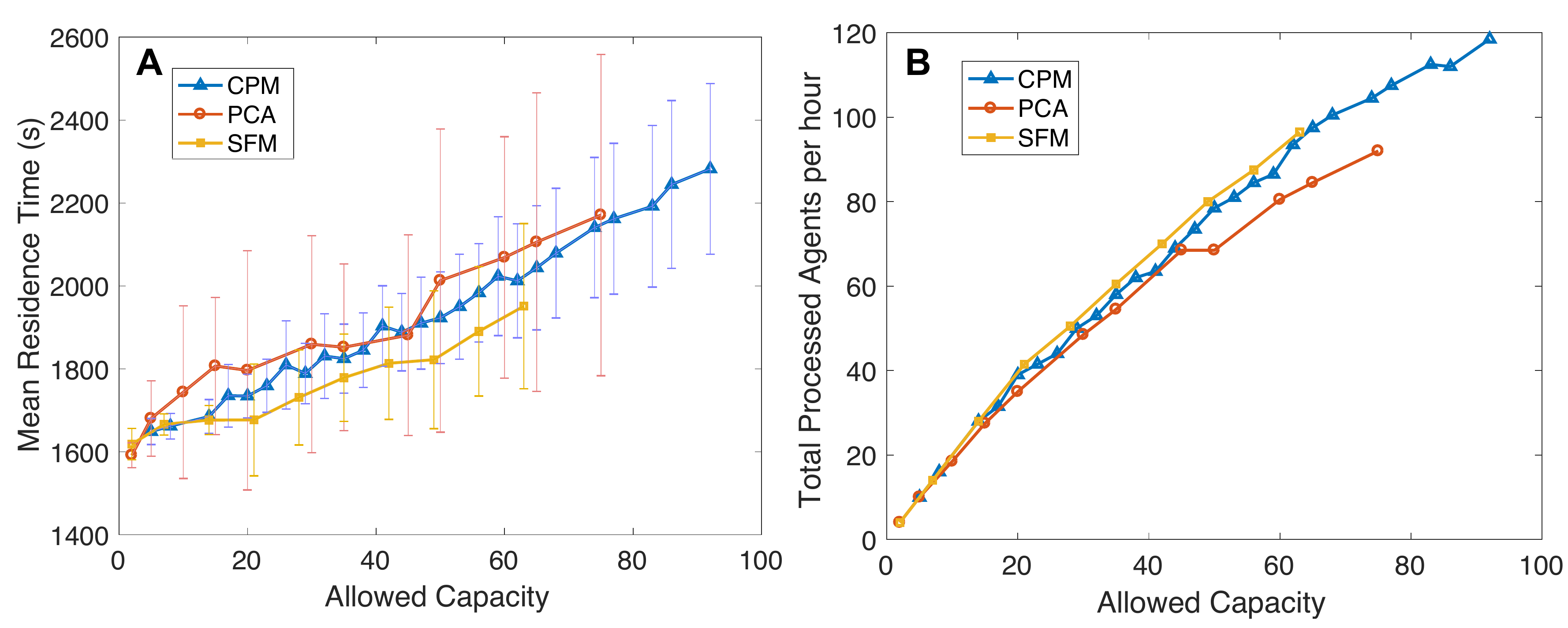}}
    \caption{A: Mean residence time of agent as a function of the occupation, for 3 different operational models. Error bar indicate one standard deviation. B: Number of processed agents per hour for the whole 2-hour simulations, also for the different operational models.}
  \label{Fig:ResTime}
\end{figure}

Also, as a macroscopic observable of the system, we study the number of agents that could be processed (i. e. fulfill the shopping list and exit the supermarket between the two hours simulated) and the mean residence time for those agents. The results are presented in Fig. \ref{Fig:ResTime}. As can be observed, both quantities increase monotonically with the allowed occupancy for the studied range of values and the supermarket set up, considering eighth checkout desks.

Also, it can be seen that different operational models display similar observables. The SFM \cite{helbing2000simulating, johansson2007specification} and PCA \cite{karamouzas2009predictive, helbing2000simulating} models are force-based models which present more limitations in terms of the maximum density they can simulate before forces get balanced (generating dead-locks) for the complex scenarios and behavior considered. This is the reason why the maximum occupancy studied with these models is lower than that simulated with the CPM described in Sec.\ref{sec:Mod}.

\subsection{Distance Analysis}
\label{Subsec:Distance}

\begin{figure}
    \centering
    \centerline{\includegraphics[width=0.5\textwidth]{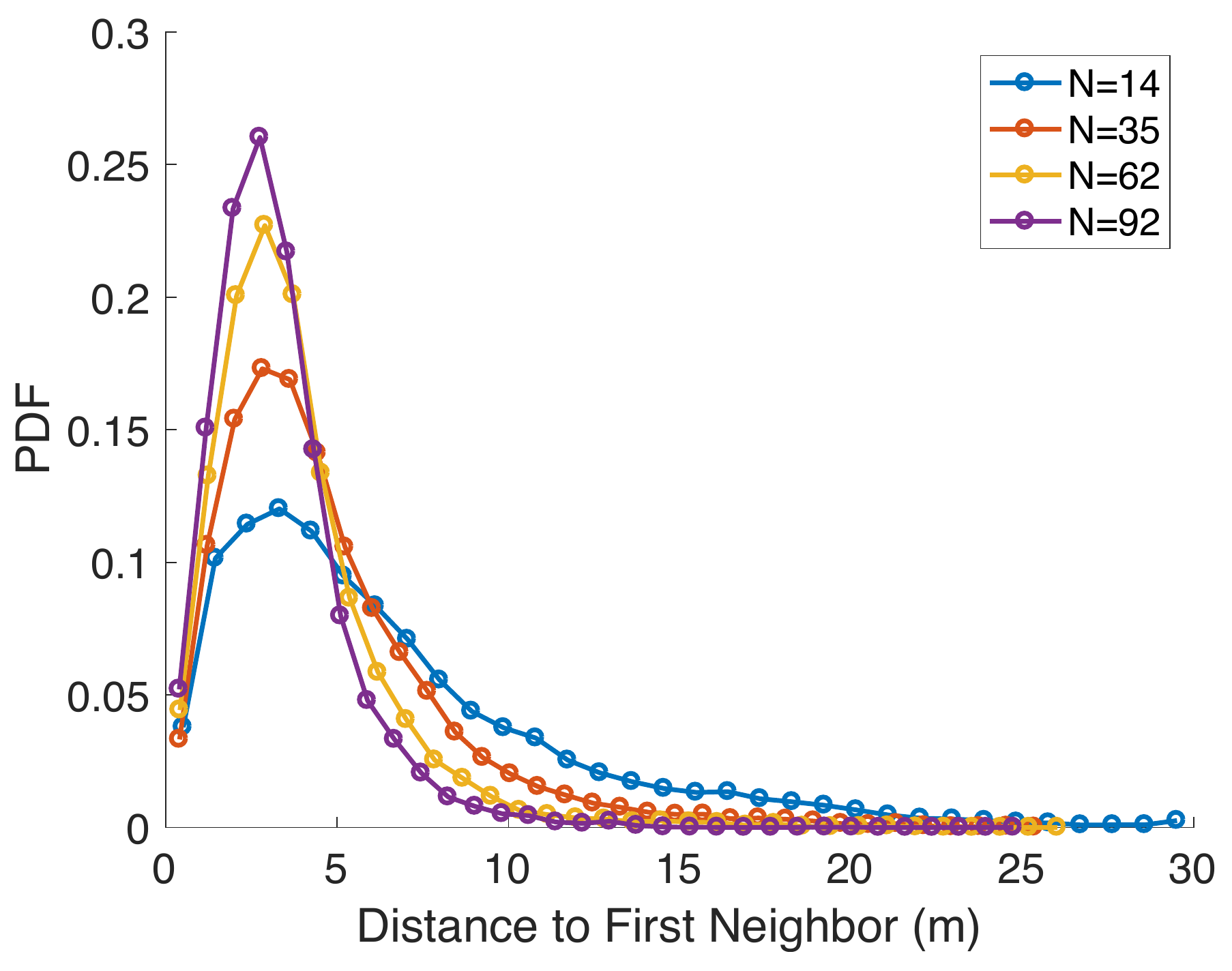}}
    \caption{Probability density function of first neighbor distances.}
  \label{Fig:FirstNeigDist}
\end{figure}

In this subsection, we characterize distance between agents during the simulations with the modified contractile particle model (CPM) for different allowed capacities. An interesting outcome is the distance to the first neighbor for each agent shown in Fig. \ref{Fig:FirstNeigDist}.

The PDF of first-neighbor distances ($d_{fn}$) shows that for lower occupations of the simulated supermarket, the probability of having the first neighbors further away than $d_{fn} \sim 5$ m is greater. On the other hand, higher occupancy values display higher probabilities of having a distance shorter than 5 m. In particular, all distributions show a maximum probable value around $d_{fn} \sim 4$ m. Moreover, the height of these probability peaks decreases for lower occupancy values.

Now we take the social distance threshold of 2 m, as discussed in Sec.\ref{sec:Int}, and calculate related probabilities of agents below this critical social distance. The first observable we calculate is the probability of having the first neighbor closer than 2 m ($P_{fn<2m}$). In other words, this is the probability of having at least one neighboring agent within 2 m. It is determined by averaging over recorded data every $\Delta t2 ~ = ~ 0.5 $ s, from minute 20 to 120 as shown in Eq.(\ref{eqProb_FN_2m})

\begin{equation}
\label{eqProb_FN_2m}
P_{fn<2m} = \frac{1}{n_{ti}} \sum_{ti=2400}^{ti=14400} \frac{n_{fn2m}}{N} 
\end{equation}
\vspace{0 pt}

where $n_{ti} = 12000 = 14400-2400$ is the data at recorded times after 20 minutes, $N$ is the occupancy and $n_{fn2m}$ is the number of particles having a first neighbor at less than 2 m. Note that if two particles $i$ and $j$ are the only particles at less than 2 m, $n_{fn2m} = 2$. Also, if $j$ is the first neighbor of $i$, not necessarily, $i$ will be the first neighbor of $j$.

The above probability ($P_{fn<2m}$) only considers if the first neighbor is closer than 2 m, but it does not take into account if there are many occurrences of neighbors at less than 2 m. For this reason, we now take into account the number of pairs having distance less than 2 m, and the corresponding probability ($P_{pair<2m}$)

\begin{equation}
\label{eqProb_Pair_2m}
P_{pair<2m} = \frac{1}{n_{ti}} \sum_{ti=2400}^{ti=14400} \frac{n_{p2m}}{[N~(N-1)]/2} 
\end{equation}
\vspace{0 pt}
where $n_{p2m}$ is the number of pairs of particles at distance closer than 2 m and $[N~(N-1)]/2$ is the total number of possible pairs, having $N$ particles in the system. In this case, if only particles $i$ and $j$ are closer than 2 m, $n_{p2m} = 1$ because one pair is counted.

\begin{figure}
    \centering
    \centerline{\includegraphics[width=1\textwidth]{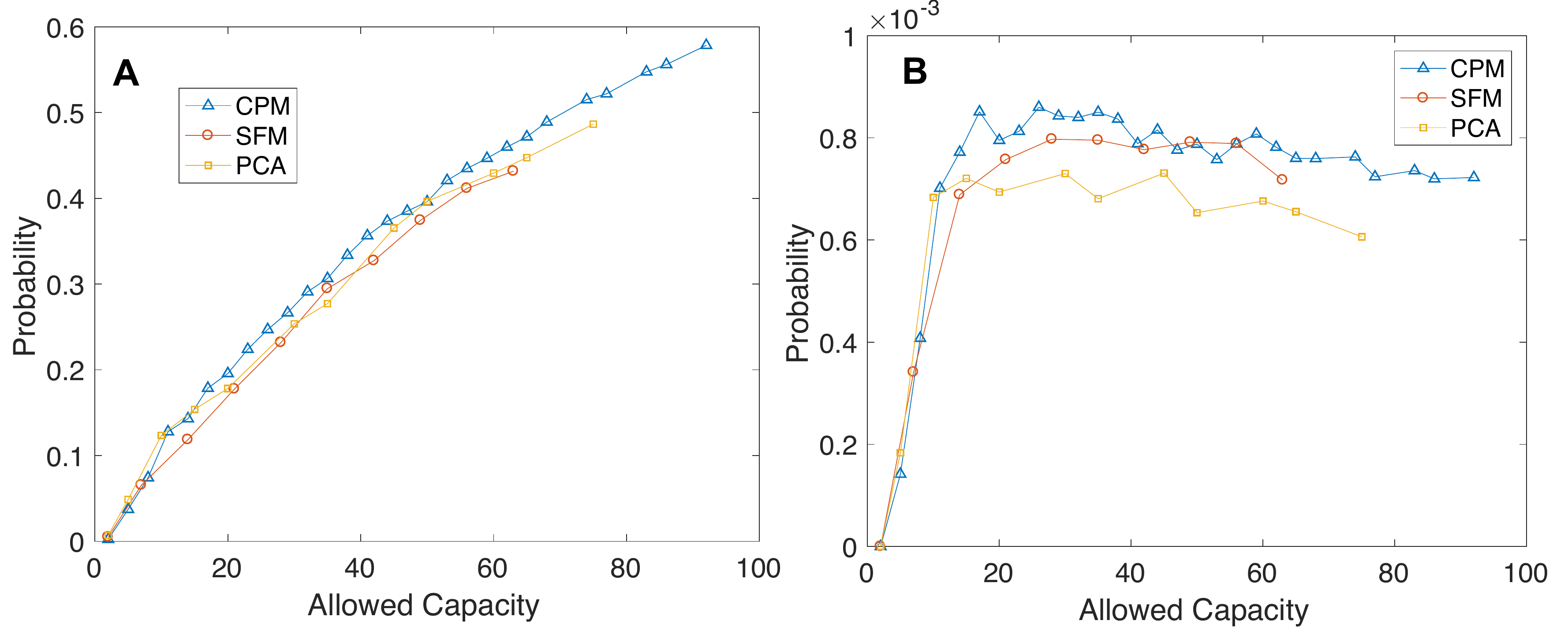}}
    \caption{A: Probability of having the first neighbor closer than 2 m (Eq.\ref{eqProb_FN_2m}). B:Probability of having any pair of agents at less than 2 m (Eq.\ref{eqProb_Pair_2m}).}
  \label{Fig:Proba}
\end{figure}

In Fig. \ref{Fig:Proba} both probabilities ( $P_{fn<2m}$ and $P_{pair<2m}$ ) are displayed for the modified contractile particle model (CPM) and also by comparison with the social force model (SFM) and predictive collision avoidance model (PCA). It can be seen that the probability of having the nearest neighbor at less than 2 m increases monotonically with the allowed capacity. However, the pair probability quickly increases for low occupancy, and after $N \sim 15$, it remains nearly constant indicating that the number of pairs $n_{p2m}$ scaled with $N$ as the number of total possible pairs ($\sim N^2$).

Furthermore, Fig. \ref{Fig:Proba} indicates that different operational models display similar macroscopic behavior regarding social distances, at least for values bellow or above 2 m.

In the above analysis, the occurrence of certain distances between simulated agents was studied, but the duration of these events was not considered explicitly. This will be done in the following subsection.

\subsection{Duration of Social Distance Events}
\label{Subsec:Duration}

Here we study the time that lasts the events when pairs of agents are found at less than 2 m (see Sec.\ref{sec:Int}). The main sources of these events are when agents are picking products at neighbor product locations or when queuing for checking out from the supermarket. If two particles $i$ and $j$ encounter at a given time, then they separate for more than 2 m, and the same particles re-encounter at a future time, it is considered as two separate events.

\begin{figure}
    \centering
    \centerline{\includegraphics[width=1\textwidth]{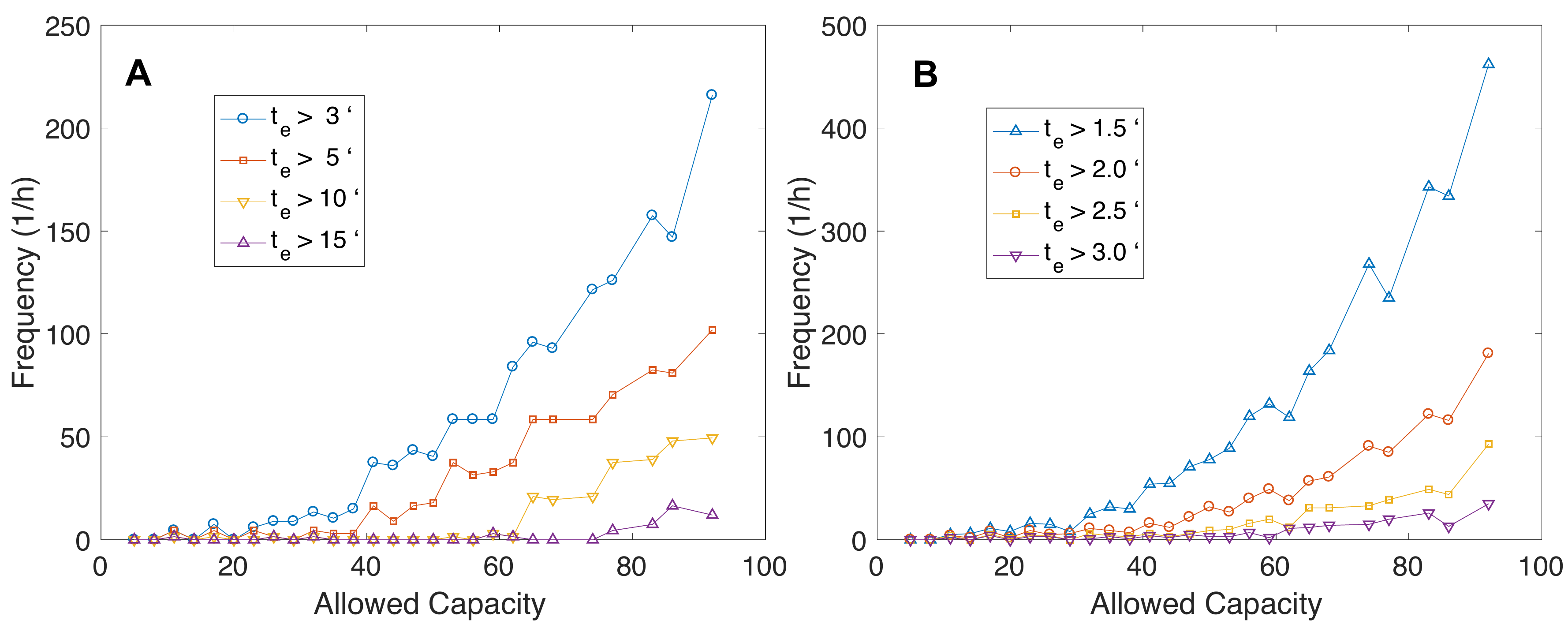}}
    \caption{A: Number of events when two agents are at distance less than 2 m during more than $t_e$ minutes registered in the first hour of simulations. B: The same measure as A but for the second hour of the simulations.}
  \label{Fig:Duration}
\end{figure}

Considering that: (a) The parameter we choose to maintain constant during each simulation is the allowed capacity $N$, and this capacity is reached at the beginning of each simulation in a very short time comparing to other processes, and (b) All agents have the same number of items in their list, and thus the required time to complete it is similar in average. The first group of $N$ agents will go to the checkout points at nearly the same time, producing the greater demand for checkout, which generates the longest queues. After that, the new agents will enter slowly as other agents exit the simulation, and thus the described behavior will relax. This dynamic lead to more queuing agents during the first hour of simulation and less during the second hour. Because of it, we analyze separately the duration of encounters occurred during the first and the second simulation hour in Fig. \ref{Fig:Duration}. The different time scales and the number of cases in its both panels confirm that the first hour is dominated by particular long lines waiting for checkout, while in the second hour (Fig. \ref{Fig:Duration} B) the duration of social distance events less than 2 m are dominated by the shorter process, i.e.: the picking time at products.

Events in the queuing line are much long-lasting because of two reasons. First, the particular process at the checkout desk takes between 2 to 4 minutes (instead of 1 to 1.5 minutes in the picking process). Second, a line with $n_l$ agents will produce that the last agents will spend about $n_l$ times $t_{co}$, which for a few agents, namely $n_l=5$, it could represent 20 minutes of waiting time at a distance of 1.5 m from another agent.

This problem of high exposure time between pairs of agents at queuing lines could be avoided if a slower ramp of inflow of agents was adopted in the start of the process, let's say something above the maximum average outflow of the system (eight agents in three minutes, i.e. $\sim$ 1 agent every 23 s). We did not adopt this in the simulations because it would take too long for simulations to reach the desired occupation $N$. But it is clear that the problem noted above at the beginning could be solved in a real operation by allowing a low flow rate of agents at the opening (of about twice the capacity of the checkout). Also, this problem would be a transient behavior only at opening, being the most part of the day operation as described in our second simulation hour.

Furthermore, Fig.8 shows that, as expected, fewer social distance events occur when the time thresholds increase. And in all cases, the number of events seems to grow quadratically with $N$.

\subsection{Social Distance Coefficient}
\label{Subsec:SDDC}

Now, looking for a criterion that determines which would be a reasonable allowed capacity in the ideal supermarket, we define the social distance coefficient ($SD_c(t_e)$), for the threshold distance of 2 m, as 

\begin{equation}
\label{eq:SDc}
SD_c(t_e)~=~\frac{2~N_e(t_e)}{N_p}
\end{equation}
\vspace{0 pt}

where $t_e$ is the minimum duration of a particular social distance event ($r_{ij} \leq 2$ m), $N_e(t_e)$ is the number of these events which last at least $t_e$, and $N_p$ is the total number of agents processed by the system in the same period of time in which $N_e$ is computed. Factor 2 is needed for taking into account the number of agents in the numerator since 2 agents ($i$ and $j$) participate in each event.

This coefficient allows us to compare how many agents have participated in social distance events of duration greater than $t_e$ concerning the number of people who have passed by the system. Thus, a value of $SD_c( t_e > 2 \text{min.})~ = ~ 1$ indicates that, on average, each agent has participated in one event of social distance less than 2 m that lasts at least 2 minutes. If $SD_c( t_e > 2 \text{min.})~ < ~ 1$, it would indicate that only a fraction of the agents have participated in such events.

\begin{figure}
    \centering
    \centerline{\includegraphics[width=1\textwidth]{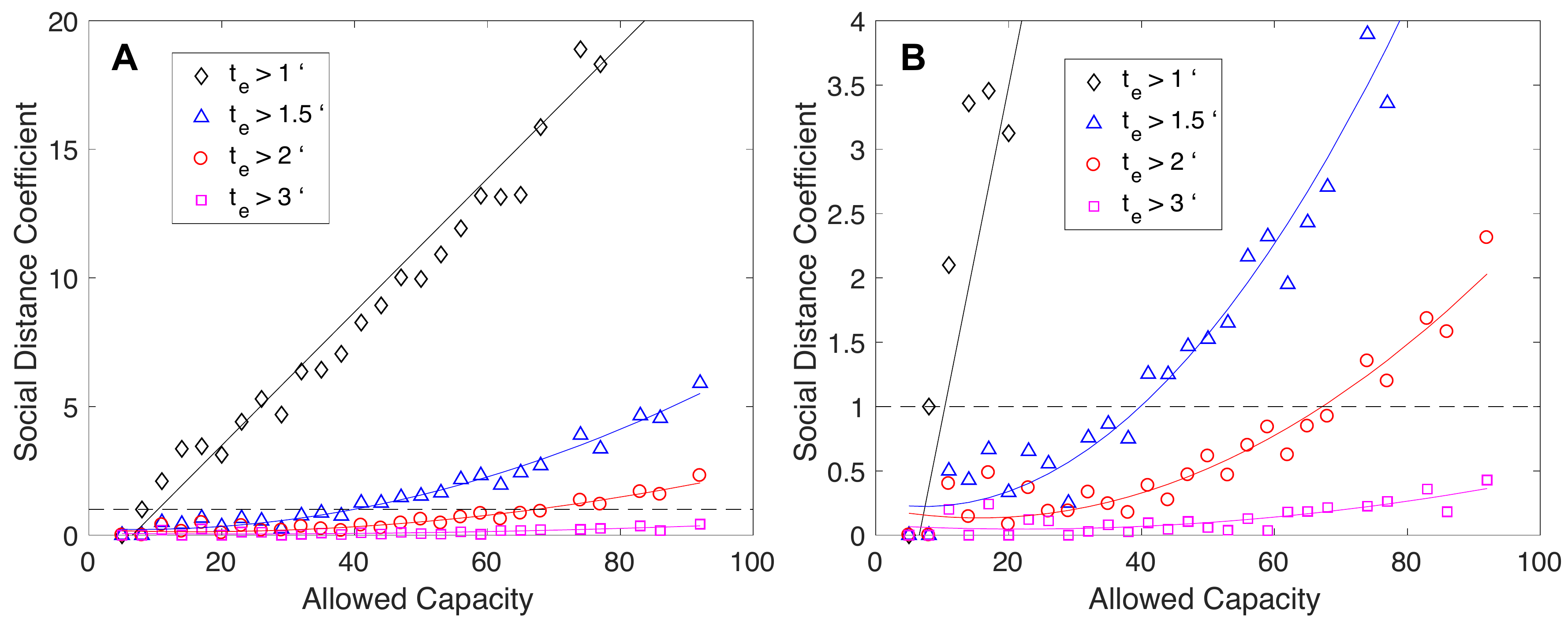}}
    \caption{A: Social distance coefficient as a function of the supermarket occupation for the second simulation hour. B: Zoom of previous figure to watch details near $SD_c \sim 1 $.}
  \label{Fig:SDcoeficient}
\end{figure}

Having established in Subsec. \ref{Subsec:Duration} that the event duration of the first simulation hour is dominated by checkout lines, we now concentrate on looking at the second hour of simulation when the impact of these lines is very low and stationary. This situation is representative of the daily operation of the supermarket, and it is shown in Fig.\ref{Fig:SDcoeficient} displaying the social distance coefficient as a function of the occupation for different event duration limits $t_e$.

First, we note in Fig. \ref{Fig:SDcoeficient} A that the curve corresponding to $t_e~>~1$ min. grows steeply with N. This could be related to the fact that the picking time range between 1 minute and 1.5 minutes, and that the products are spaced by 1 m, so if two agents must go simultaneously to the same product or the first or second nearest product, they could generate a 2 m social event lasting at most 1.5 minutes, in particular many events lasting more than 1 minute would occur. Also, the social distance coefficient seems to follow a linear relation with $N$ for this particular time limit $t_e$.

A change of regime can be observed for $t_e~>~ 1.5$ min. in which curves are more similar between themselves for the different $t_e$ presented, and they follow a quadratic relation with $N$. Because the maximum picking time is 1.5 minutes, this is the maximum possible overlapping time for two agents in neighbor (or the same) products. Greater time events will arise when more than two agents are waiting for the same or near products, as is in the cases of products near any of the short lines for checking out.

Finally, we could use Fig. \ref{Fig:SDcoeficient} B, as a guide for deciding on the allowed occupancy. If based on epidemiological knowledge or criteria, it was decided that it would be acceptable to allow that all agents participate once in a 2-m social event lasting at most 1 minute, then the allowed occupation would be very small, $N~\sim~10$. Alternatively, if events up to 1.5 minutes were accepted, then the allowed occupation would be $N~=~40$. In the case of $t_e ~=~ 2$ min., the capacity could rise to $N~=~70$. Also, it could be established that even for $N~=~90$ the events of the social distance of 2 m, lasting more than 3 minutes, would only affect the 40\% of the processed agents.

Of course, the same Fig. \ref{Fig:SDcoeficient} B could be used for finding another allowed occupancy if the criterion would consider that, for example, only 25\% of the agents can participate in the analyzed events.

\section{Conclusions}
\label{sec:Con}

In this work, we investigate and characterize the social distance realization in an everyday pedestrian system by simulating the dynamics of an ideal supermarket. Many sources of complexity were successfully taken into account with a multi-level model, which allows us to simulate not only translation but also more complex behaviors like waiting times when picking for particular products and queuing for checkout points.

The main process that keeps pedestrians near one to another is the queuing lines for checkout. Therefore, advice for the operation would be to keep these lines as small as possible either by increasing the number of checkout points or by decreasing the occupancy.

Different operational models, display similar macroscopic observables regarding social distances at values greater than 2 m indicating that the results are robust with respect to microscopic collision avoidance resolution and also suggesting that the simulated paths of the particles are more influenced by the geometry, shopping list, and time-consuming process, than by the particular avoidance mechanism. However, first-order models as the CPM presented in \cite{baglietto2011continuous} and Sec.\ref{sec:Mod} seem more suitable for simulation of high populated scenarios with complex behavioral agents.

Taking a social distance threshold of 2 m \cite{dbouk2020coughing}, probabilities and duration of such events were studied. The social distance coefficient was defined as an indicator of the fraction of the population passing by the system that is involved in one or many of these events lasting at least a certain time threshold $t_e$.

The same analysis can be done for a different set of parameters and, of course, for other pedestrian facilities, being other specific supermarkets or any different systems (Transport, Entertainment, etc.). Of course, existing facilities can be monitored with measurement methods \cite{pouw2020monitoring} providing high-quality trajectory data. This kind of data could also be interpreted in terms of the analysis performed in the present work.

By setting a distance and a time threshold, and what fraction of the population would be tolerable to be exposed to these conditions, the maximum occupancy can be established in any particular system by performing simulations and using the observables defined in this work. Therefore, this analysis could be used for determining which is the ideal occupation of a pedestrian facility, keeping low contagion risk, and maximizing the number of users per unit of time.

\section*{Acknowledgements}

The authors acknowledge the information and data provided by the Argentinean Supermarket chain "La Anónima". 
This work was founded by project PID2015-003 (ANPCyT) Argentina.


\bibliographystyle{unsrt}
\bibliography{SocialDistanceSupermarket.bib}

\begin{thebibliography}{10}

\bibitem{who2020coronavirus}
World~Health Organization.
\newblock {Coronavirus disease (COVID-19) advice for the public}.
\newblock
  \url{https://www.who.int/emergencies/diseases/novel-coronavirus-2019/advice-for-public}.
\newblock Last updated: 4 June 2020.

\bibitem{rathinakumar2020microscopic}
Krithika Rathinakumar and Annalisa Quaini.
\newblock A microscopic approach to study the onset of a highly infectious
  disease spreading.
\newblock {\em arXiv preprint arXiv:2004.09554}, 2020.

\bibitem{harweg2020agent}
Thomas Harweg, Daniel Bachmann, and Frank Weichert.
\newblock Agent-based simulation of pedestrian dynamics for exposure time
  estimation in epidemic risk assessment.
\newblock {\em arXiv preprint arXiv:2007.04138}, 2020.

\bibitem{pouw2020monitoring}
Caspar~AS Pouw, Federico Toschi, Frank van Schadewijk, and Alessandro Corbetta.
\newblock Monitoring physical distancing for crowd management: real-time
  trajectory and group analysis.
\newblock {\em arXiv preprint arXiv:2007.06962}, 2020.

\bibitem{dbouk2020coughing}
Talib Dbouk and Dimitris Drikakis.
\newblock On coughing and airborne droplet transmission to humans.
\newblock {\em Physics of Fluids}, 32(5):053310, 2020.

\bibitem{hoogendoorn2004pedestrian}
Serge~P Hoogendoorn and Piet~HL Bovy.
\newblock Pedestrian route-choice and activity scheduling theory and models.
\newblock {\em Transportation Research Part B: Methodological}, 38(2):169--190,
  2004.

\bibitem{yao2010path}
Junfeng Yao, Chao Lin, Xiaobiao Xie, Andy~JuAn Wang, and Chih-Cheng Hung.
\newblock Path planning for virtual human motion using improved a* star
  algorithm.
\newblock In {\em 2010 Seventh international conference on information
  technology: new generations}, pages 1154--1158. IEEE, 2010.

\bibitem{baglietto2011continuous}
Gabriel Baglietto and Daniel~R Parisi.
\newblock Continuous-space automaton model for pedestrian dynamics.
\newblock {\em Physical Review E}, 83(5):056117, 2011.

\bibitem{martin2020pedestrian}
Rafael Martin and Daniel Parisi.
\newblock Pedestrian collision avoidance with a local dynamic goal.
\newblock {\em Collective Dynamics}, 5:324--331, 2020.

\bibitem{helbing2000simulating}
Dirk Helbing, Ill{\'e}s Farkas, and Tamas Vicsek.
\newblock Simulating dynamical features of escape panic.
\newblock {\em Nature}, 407(6803):487--490, 2000.

\bibitem{johansson2007specification}
Anders Johansson, Dirk Helbing, and Pradyumn~K Shukla.
\newblock Specification of the social force pedestrian model by evolutionary
  adjustment to video tracking data.
\newblock {\em Advances in complex systems}, 10(supp02):271--288, 2007.

\bibitem{karamouzas2009predictive}
Ioannis Karamouzas, Peter Heil, Pascal Van~Beek, and Mark~H Overmars.
\newblock A predictive collision avoidance model for pedestrian simulation.
\newblock In {\em International workshop on motion in games}, pages 41--52.
  Springer, 2009.

\end{thebibliography}

\end{document}